\documentclass[aps,prl,twocolumn,superscriptaddress,showpacs,amsmath,amssymb]{revtex4}

\usepackage{graphicx}% Include figure files
\usepackage{bm}% bold math
\usepackage{amssymb}
\usepackage{hyperref}

\begin{document}

\title{Weak Measurements of Light Chirality with a Plasmonic Slit}

\author{Y. Gorodetski}
\affiliation{ISIS, Universit\'{e} de Strasbourg and CNRS (UMR 7006), 8 all\'{e}e Gaspard Monge, 67000 Strasbourg, France}

\author{K. Y. Bliokh}
\affiliation{Advanced Science Institute, RIKEN, Wako-shi, Saitama 351-0198, Japan}
\affiliation{A. Usikov Institute of Radiophysics and Electronics, NASU, Kharkov 61085, Ukraine}

\author{B. Stein}
\affiliation{ISIS, Universit\'{e} de Strasbourg and CNRS (UMR 7006), 8 all\'{e}e Gaspard Monge, 67000 Strasbourg, France}

\author{C. Genet}
\affiliation{ISIS, Universit\'{e} de Strasbourg and CNRS (UMR 7006), 8 all\'{e}e Gaspard Monge, 67000 Strasbourg, France}

\author{N. Shitrit}
\affiliation{Micro and Nanooptics Laboratory, Faculty of Mechanical Engineering and Russell Berrie Nanotechnology Institute, Technion--Israel Institute of Technology, Haifa 32000, Israel}

\author{V. Kleiner}
\affiliation{Micro and Nanooptics Laboratory, Faculty of Mechanical Engineering and Russell Berrie Nanotechnology Institute, Technion--Israel Institute of Technology, Haifa 32000, Israel}

\author{E. Hasman}
\affiliation{Micro and Nanooptics Laboratory, Faculty of Mechanical Engineering and Russell Berrie Nanotechnology Institute, Technion--Israel Institute of Technology, Haifa 32000, Israel}

\author{T. W. Ebbesen}
\affiliation{ISIS, Universit\'{e} de Strasbourg and CNRS (UMR 7006), 8 all\'{e}e Gaspard Monge, 67000 Strasbourg, France}

\begin{abstract}
We examine, both experimentally and theoretically, an interaction of tightly focused polarized light with a slit on a metal surface supporting plasmon-polariton modes. Remarkably, this simple system can be highly sensitive to the polarization of the incident light and offers a perfect quantum-weak-measurement tool with a built-in post-selection in the plasmon-polariton mode. We observe the plasmonic spin Hall effect in both coordinate and momentum spaces which is interpreted as weak measurements of the helicity of light with real and imaginary weak values determined by the input polarization. Our experiment combines advantages of (i) quantum weak measurements, (ii) near-field plasmonic systems, and (iii) high-numerical aperture microscopy in employing spin-orbit interaction of light and probing light chirality.
\end{abstract}


\maketitle

\textit{Introduction.---}
%%%%%%%%%%%%%%%%%%%%%%%%%%%%%%%%%%%%%%%%%%%%%%%%%%%%%%%%%%%
Polarization-dependent \textit{transverse shifts} of spatially-confined optical beams, also known as the \textit{spin Hall effect of light} (SHEL), has become a topic of an intensive research since pioneering studies by Fedorov and Imbert \cite{Fedorov,Imbert} and other important early works \cite{Schilling,Player,Dutriaux}. The SHEL manifests itself in opposite out-of-plane displacements of the trajectories of right- and left-hand circularly polarized beams reflected or refracted by a plane interface. Fundamentally, this subwavelength phenomenon stems from a \textit{spin-orbit interaction} (SOI) of light, i.e., a weak coupling of photon spin (helicity or chirality) and the trajectory of light propagation \cite{LZ,Bliokh et al,OMN}. During the past few years, interest in spin-dependent transverse shifts has grown intensively \cite{OMN,BB1,BB2,HK,Aiello,Qin} (for a review, see \cite{Review}), motivated by the rapid development of spintronics and nano-optics. After 50 years of highly controversial studies, an accurate theoretical description of the SHEL at a plane dielectric interface was formulated \cite{BB1,BB2} and completely verified in a remarkable experiment by Hosten and Kwiat \cite{HK} (see also \cite{Aiello,Qin,Review}).

Importantly, the experiments of \cite{HK,Qin} achieved incredible angstrom accuracy in determination of the SHEL shift owing to the method of \textit{quantum weak measurements} \cite{AAV,Duck,Jozsa} (for reviews, see \cite{Nori}). It was shown that purely classical interaction of a transversely-confined polarized optical beam with a plane interface can be interpreted as a quantum weak measurement of the photon spin (helicity) by the transverse profile of the beam, which is described by the optical SOI Hamiltonian. Owing to this, employing almost orthogonal polarizers before and after the interface (i.e., pre-selection and post-selection of the spin states), one can enormously magnify the observed beam shift from the subwavelength to beam-width scale. This measured shift represents the actual SHEL shift multiplied by the \textit{weak value} of photon helicity which can take large complex values. In this manner, \textit{real} weak values correspond to \textit{spatial displacements} of the beam, whereas \textit{imaginary} weak values correspond to \textit{angular deflections} of the beam (i.e., shifts in the \textit{momentum} space) \cite{BB2,HK,Aiello,Qin,Review}. So far, weak-measurement SHEL experiments used only imaginary weak values and angular shifts because they result in much higher beam shifts in the far-field.

Alongside classical-optics far-field systems, the subwavelength nature of the SHEL makes it highly relevant and attractive for near-field optics \cite{Focusing1}, high-numerical-aperture (NA) microscopy \cite{Focusing2}, and, particularly, plasmonics \cite{Plasmonic1,Plasmonic2}. In these areas, the optical SOIs dramatically modify distributions of near fields and offer promising applications. In the present Letter, we combine the fundamental advantages of (i) \textit{quantum weak measurements}, (ii) \textit{near-field plasmonic system}, and (iii) \textit{high-NA microscopy}. We show that coupling of a tightly focused optical beam to surface plasmon polaritons (SPPs) offers a natural weak-measurement tool with a built-in post-selection provided by the fixed linear polarization of the SPP mode \cite{Ebbesen}. We use slightly tilted linear and slightly elliptical input polarizations of light which provide \textit{both imaginary and real} weak values of spin and measure \textit{both spatial and angular transverse shifts} in the SPP beam launched by a single slit. These measurements are performed on a leakage radiation microscope \cite{LRM} which allows visualization of a plasmonic beam in real and momentum (Fourier) spaces.

\textit{Experiment and weak-measurement model.---}
%%%%%%%%%%%%%%%%%%%%%%%%%%%%%%%%%%%%%%%%%%%%%%%%%%%%%%%%%%%
The experimental setup is schematically shown in Figure 1. We used a sample consisted of a glass wafer, coated with a thin layer of gold (the thickness is about 70nm). A straight 100nm-wide straight slit was milled in the metal using focused ion beam. The slit was illuminated by a focused laser beam (785nm) prepared using objectives with the numerical apertures ${\rm NA}=0.45$ and ${\rm NA}=0.6$. Upon interaction with the slit, the incident optical beam is partially scattered into two SPP beams propagating along the gold layer orthogonally to the slit (Fig. 1). The SPP beams were observed using leakage signal collected via an immersion objective attached to the back side of the sample. More detailed description of the standard leakage radiation microscope setup can be found elsewhere \cite{LRM}. Note that the incident beam was focused to a plane behind the gold layer, so that the actual focal spot occurs in the secondary, SPP beams (see Fig. 2). Using a polarizer with rotating quarter-wave and half-wave plates at the input of the system, we were able to produce an arbitrary polarization state of the incident light (Fig. 1).

Remarkably, we observed extraordinary asymmetric deformations in the SPP beams when the input polarization of light was just slightly off from being parallel to the slit, as shown in Fig. 2. For instance, a tiny rotation of the quarter-wave plate (producing a slightly elliptical polarization) caused a strong \textit{angular deviation} of the beams, whereas a tiny rotation of the half-wave plate (slightly tilted linear polarization) resulted in a well pronounced \textit{spatial displacement} of the focal spot. These anomalous SPP beam shifts represent \textit{plasmonic SHEL} and can be associated with ``quantum weak measurements'' \cite{AAV,Duck,Jozsa,Nori} of the incident light helicity via spin-orbit coupling induced by the light-to-SPP transformation at the slit. First, we interpret the results within a simple ``quantum weak measurements'' model of the SOI of light, and afterwards will give a complete wave description of the problem.
%
%%%%%%%%%%%%%%%%%%%%%%%%%%%%%%%%%%%%%%%%%%%%%%%%%%%%%%%%%%%
\begin{figure}[t]
\includegraphics[width=8cm, keepaspectratio]{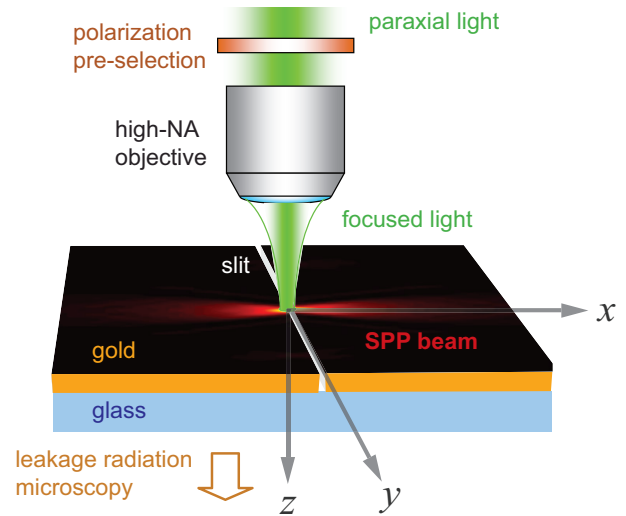}
\caption{(color online). Conceptual scheme of the experimental setup. The objective is used to focus the incident light with pre-selected polarization on a thin layer of gold with a straight slit along the $y$ axis. The slit generates SPP beams propagating in the $\pm x$ directions, and the leakage SPP signal is collected by an immersion objective.} 
\label{fig1}
\end{figure}
%%%%%%%%%%%%%%%%%%%%%%%%%%%%%%%%%%%%%%%%%%%%%%%%%%%%%%%%%%%
%

Let the incident light propagate along the $z$-axis ($z=0$ represents the gold surface), the slit be parallel to the $y$-axis, whereas the SPP beams propagate in the $\pm x$ directions (from now on we consider only the $+x$ beam), Fig.~1. According to the ``weak measurements'' formalism, the external transverse spatial profile of the beam, $\Phi(y)$, and its internal polarization state, $\left| \Psi \right\rangle$, represent ``classical measuring subsystem'' and ``quantum measured subsystem'', respectively \cite{Duck}. For simplicity, let the $y$-distribution of the incident light be Gaussian in the focal plane:
\begin{equation}\label{eqn:1}
\Phi_{in} \propto \exp \left[ - y^2/w_{0}^{2} \right]~, 
\end{equation}
where $w_0 \sim \left( k {\rm NA} \right)^{-1}$ is the beam waist and $k$ is the wavenumber of light. At the same time, using the basis of linear polarizations $\left| X \right\rangle$ and $\left| Y \right\rangle$ along the corresponding axes, and the spin basis of right- and left-hand circular polarizations, $\left| R \right\rangle$ and $\left| L \right\rangle$, the \textit{pre-selected} input polarization of light is chosen to be \textit{almost $y$-linear}:
\begin{equation}\label{eqn:2}
\left| \Psi_{in} \right\rangle \simeq \left| Y \right\rangle - \varepsilon \left| X \right\rangle  = \frac{{\left(- i - \varepsilon \right)\,\left| R \right\rangle  + \left(i - \varepsilon \right)\,\left| L \right\rangle }}{\sqrt{2}}~, 
\end{equation}
Here $\varepsilon$ ($|\varepsilon|\ll 1$) is a complex number, with \textit{real} and \textit{imaginary} $\varepsilon$ corresponding to the \textit{slightly tilted linear} and \textit{slightly elliptical} polarizations, respectively (see Fig.~2). In the spin basis, states $\left| R \right\rangle = \left(1,0\right)^T$ and $\left| L \right\rangle = \left(0,1\right)^T$ are the eigenvectors of the photon \textit{helicity} operator 
$\hat \sigma _3  = {\rm diag} \left( 1, - 1 \right)$ \cite{Review}.

Interaction of light with the slit and transformation to SPPs is similar to the beam refraction at a plane interface, and the geometric-phase difference between the constituent plane-wave components produce the SOI of light \cite{HK,Review,Remark I}. One can show \cite{Plasmonic1} that the geometric-phase factor for the generated SPP waves with different $\bm{k}$-vectors is $\exp \left( {i\hat \sigma _3 k_y /k} \right)$ for $k_y \ll k$ (cf. \cite{HK,Review}), which implies the dimensionless SOI Hamiltonian
\begin{equation}\label{eqn:3}
\hat H_{{\mathop{\rm SOI}\nolimits} }  \simeq  - \lambdabar \hat \sigma _3 k_y~, 
\end{equation}
where $\lambdabar=k^{-1}$ is the SHEL shift playing role of the coupling constant \cite{HK}. Employing the weak-measurement interpretation \cite{HK,Duck}, the transverse profile of the beam, Gaussian-distributed in $y$ and $k_y$ ``weakly measures'' its helicity $\hat\sigma_3$ multiplied by the SHEL constant $\lambdabar$ at the moment of interaction with the interface (slit).

Since the $y$-component of the incident electric wave field cannot interact with plasmons via the slit, this naturally defines the \textit{post-selected} polarization state to be perpendicular to the slit:
\begin{equation}\label{eqn:4}
\left| {\Psi _{out} } \right\rangle  = \left| X \right\rangle  = \frac{{\left| L \right\rangle  + \left| R \right\rangle }}{{\sqrt 2 }}~.
\end{equation}
In fact, this $\left| X \right\rangle$ state in the \textit{local} coordinate frame attached to the direction of propagation of the beam corresponds to the \textit{$z$-component} of the SPP beam propagating along the $x$-axis.

In terms of weak measurements, the input and output polarization states (2) and (4) determine the \textit{weak value} $\sigma_w$ of the photon helicity \cite{HK,AAV,Duck,Jozsa,Nori}:
\begin{equation}\label{eqn:5}
\sigma_w  = \frac{{\left\langle {\Psi _{out} } \right|\left. {\,\hat \sigma _3 \,} \right|\left. {\Psi _{in} } \right\rangle }}
{{\left\langle {{\Psi _{out} }}\left.\right|{{\Psi _{in} }} \right\rangle }} = \frac{i}{\varepsilon }~. 
\end{equation}
Remarkably, this weak value is complex and large, $|\sigma_w|\gg 1$, although the photon helicity eigenvalues are $\sigma=\pm 1$. It is seen from Eq.~(5) that elliptical and linear tilted pre-selected polarizations yield real and imaginary $\sigma_w$, respectively. Weak measurement of helicity, Eq. (5), results in the transverse shift of the ``measuring subsystem'', i.e., of the transverse beam profile \cite{Duck}:
\begin{equation}\label{eqn:6}
\Phi_{out} \propto \exp\left[ { - \left( {y - \Delta } \right)^2 }/{w_0^2 } \right]~,~~\Delta \simeq -\lambdabar \sigma_w~.
\end{equation}
Thus, the output SPP beam undergoes \textit{complex transverse shift} (6) equaling to the SHEL shift $\lambdabar$ multiplied by the helicity weak value (5). In this manner, real and imaginary parts of $\sigma_w$ produce \textit{coordinate shift} (displacement) and \textit{momentum shift} (deflection) of the beam profile \cite{BB2,HK,Aiello,Qin,Review,Jozsa}:
\begin{equation}\label{eqn:7}
\left\langle y \right\rangle  = {\rm Re} \Delta~,~~\left\langle {k_y } \right\rangle  = 2w_0^{-2}\,{\rm Im} \Delta~.
\end{equation}
Note that the weak-measurement approximation fails in the vicinity of $\varepsilon=0$ and is applicable at $\lambdabar/w_0 \ll \varepsilon \ll 1$ \cite{Duck}.

Compare now the above weak-measurement model, based on the SOI of light, with the experimental plasmonic results presented in Fig. 2. Both coordinate and momentum shifts are clearly visible in the SPP fields in real (a,b) and Fourier (c,d) spaces for two types of the pre-selected polarization. However, the observed coordinate and momentum shifts of the SPP beams turn out to be \textit{swapped} as compared to the model (5)--(7): Real $\varepsilon$ (tilted linear polarization) causes coordinate shift $\left\langle y \right\rangle \neq 0$, whereas imaginary $\varepsilon$ (elliptical polarization) induces momentum shift $\left\langle k_y \right\rangle \neq 0$. As we show below, the accurate description of the light-SPP coupling appends an additional $\pi/2$ phase factor that interchanges coordinate and momentum shifts with respect to the real or imaginary nature of $\varepsilon$. These swapped relations with respect to usual SHEL shifts have to be understood as an inherent feature of the \textit{plasmonic SHEL}.
%
%%%%%%%%%%%%%%%%%%%%%%%%%%%%%%%%%%%%%%%%%%%%%%%%%%%%%%%%%%%
\begin{figure}[t]
\includegraphics[width=8.5cm, keepaspectratio]{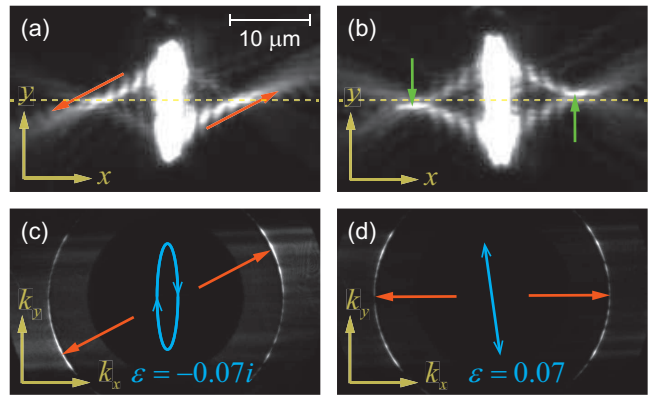}
\caption{(color online). Typical SPP field distributions in real and Fourier spaces (see also \cite{SOM}). The mean momenta $\left\langle k_y \right\rangle$ and transverse coordinates $\left\langle y \right\rangle$ of the beams are schematically indicated by orange and green arrows, respectively, whereas pre-selected polarizations of the incident light are shown in cyan. In the case of elliptical polarization (imaginary $\varepsilon$), panels (a) and (c), the SPP beams demonstrate angular SHEL deviation which is seen as the momentum shift in the Fourier space. For tilted linear polarization (real $\varepsilon$), panels (b) and (d), the SPP beams exhibit coordinate SHEL shift which is seen only in real space.} 
\label{fig2}
\end{figure}
%%%%%%%%%%%%%%%%%%%%%%%%%%%%%%%%%%%%%%%%%%%%%%%%%%%%%%%%%%%
%

\textit{Complete wave theory.---}
%%%%%%%%%%%%%%%%%%%%%%%%%%%%%%%%%%%%%%%%%%%%%%%%%%%%%%%%%%%
The detailed wave picture of light evolution in the system can be divided into three stages: (i) focusing of the initial paraxial polarized field by a high-NA objective; (ii) interaction of the focused field with the slit; and (iii) generation and propagation of the SPP beams.

First, approximating the incident light by a single plane wave with the complex electric-field amplitude ${\bf E}_0 \propto \left(- \varepsilon ,1,0\right)^T$ (in the Cartesian basis), the high-NA focusing is described by the Debye-Wolf approach \cite{Focusing2,RW}. It implies that the transverse electric field of the wave is parallel-transported along each geometrical-optics ray, refracted by the lens, without change of the polarization state in the local coordinates. Marking the rays by spherical angles $\left( {\theta ,\phi } \right)$ indicating the direction of the wave vectors ${\bf k} = k\left( {\cos \theta ,\sin \theta \cos \phi ,\sin \theta \sin \phi } \right)$, the plane-wave spectrum of the focused field, ${\bf \tilde E}_{{\mathop{\rm lens}\nolimits} } \left( {\bf k} \right) \equiv {\bf \tilde E}_{{\mathop{\rm lens}\nolimits} } \left( {\theta ,\phi } \right)$, is given by geometric rotational transformation \cite{Focusing2}:
\begin{equation}\label{eqn:8}
{\bf \tilde E}_{\rm lens}\!\left( {\theta ,\phi } \right) \propto \sqrt {\cos \theta }\;\hat T_{\rm lens}\!\left( {\theta ,\phi } \right){\bf E}_0~.
\end{equation}
Here $\hat T_{\rm lens} = \hat R_z \left( { - \phi } \right)\hat R_y \left( { - \theta } \right)\hat R_z \left( \phi  \right)$, with $\hat R_a \left( \alpha  \right)$ denoting the SO(3) matrix operator of rotation about the $a$-axis by the angle $\alpha$, and $\sqrt{\cos\theta}$ is the apodization factor which provides conservation of the energy flow \cite{RW}. The real-space focused field is given by the Fourier-type integral over all plane waves ${\bf \tilde E}_{\rm lens} \left( {\bf k} \right)$, but the coupling with the SPPs is described in the momentum representation, and we perform this integration afterwards.

Second, the focused field (8) interacts with the slit. In doing so, only $x$ and $z$ components of the field can excite SPPs, whereas the $y$ component, parallel to the slit, does not take part in the interaction. In other words, the slit acts as a polarizer which cuts out the $y$ component of the field. This is described by the \textit{projection} of the field ${\bf \tilde E}_{\rm lens}$ onto the $(x,z)$ plane: 
\begin{equation}\label{eqn:9}
{\bf \tilde E}_{\rm slit} = \hat P_y \,{\bf \tilde E}_{\rm lens}~,~~
\end{equation}
where $\hat P_y  = {\rm diag} \left( {1,0,1} \right)$ is the projection operator.

Finally, the field ${\bf \tilde E}_{\rm slit}$ can be considered as the source for the SPPs. Excited SPP field can be decomposed into plane waves $\tilde {\bm{\mathcal E}}_p$ which are characterized by the real wave vectors ${\bf k}_p  = \left({k_{p_x},k_{p_y},0}\right)$ (dissipation is neglected hereafter), exponential decay away from the metal surface, $\exp \left( {\kappa z} \right)$ at $z<0$, and the complex electric-field amplitudes $\tilde {\bm{\mathcal E}}_p  = \tilde {\mathcal E}_{p\bot}{\bf e}_z + \tilde{\mathcal E}_{p\parallel}\,{\bf k}_p/k_p$. From Maxwell equations it follows that the longitudinal and transverse field components are related as $\tilde {\mathcal E}_{p\parallel} = -i\chi \tilde{\mathcal E}_{p\bot}$, $\chi  = \kappa /k_p$ \cite{Bradberry}. Excitation of the SPPs by focused light via the slit is determined by: (i) the \textit{phase matching condition} that provides transformation of the wave momenta ${\bf k} \to {\bf k}_p$ and (ii) the \textit{coupling efficiency}, $\gamma$, which we model by the inner product of the electric-field amplitudes of the incident light and SPPs: $\gamma  \propto \tilde {\bm{\mathcal E}}_p^* \cdot {\bf \tilde E}_{\rm slit}$ \cite{Remark II}. Since the system is homogeneous in the $y$-direction, the corresponding momentum component must be conserved. Taking into account that the SPP wave number $k_p>k$ is fixed (which determines the SPP circle in the Fourier space, Fig.~2), the phase matching condition can be written as
\begin{equation}\label{eqn:10}
k_{p_y}  = k_y~~,~~~k_{p_x}  = \sqrt {k_p^2  - k_y^2 }~,
\end{equation}
where $k_y  = k\,\sin{\theta} \sin{\phi}$. From here and equations above, the SPP plane-wave amplitude is
\begin{eqnarray}\label{eqn:11}
\tilde {\bm{\mathcal E}}_p\!\propto\!\left( { - i\chi \sqrt {1 - \left( {\frac{{k_y }}{{k_p }}} \right)^2 } , - i\chi \frac{{k_y }}{{k_p }},1} \right)^{\!T}.
\end{eqnarray}
Taking into account the coupling efficiency $\gamma$, the resulting SPP field in the momentum and coordinate representations reads
\begin{eqnarray}
\label{eqn:12}
{\bf \tilde E}_p\!\propto\!\left( {\tilde {\bm{\mathcal E}}_p^*\!\cdot\!{\bf \tilde E}_{\rm slit}} \right)\!\tilde {\bm{\mathcal E}}_p,~
{\bf E}_p\!\propto\!\int\limits_0^{2\pi }\!{\int\limits_0^{\theta_{\rm c}} {\bf \tilde E}_pe^{i{\bf k}_{p}\cdot{\bf r}}}\!\sin{\!\theta} d\theta d\phi.
\end{eqnarray}
Here the second Eq.~(12) is the Fourier (Debye) integral over all incoming plane waves, $\theta_{\rm c} = \sin^{-1}\!\left({\rm NA}\right)$ is the aperture angle of the focusing objective, and we consider only the $(x,y)$-distribution of the SPP field, omitting the common $z$-dependence.

Equations (8)--(12) completely describe our system starting from the incident paraxial field ${\bf E}_0$ and yielding the output SPP field distribution ${\bf E}_p$ on the gold layer. We have verified numerically that these equations yield SPP distributions and shifts corresponding to experimental Figure 2 \cite{SOM}.

In fact, for our experiment one can use the first post-paraxial approximation, $\sin{\theta} \simeq \theta$ and $\cos{\theta} \simeq 1$, which significantly simplifies equations and allows analytical solution. Keeping only terms linear in $\theta$ and $\varepsilon$, but neglecting $\varepsilon\theta$-order terms, Eqs.~(8)--(12) yield
\begin{equation}\label{eqn:13}
%{\bf \tilde E}_{\rm lens} \propto \left({-\varepsilon,1,- \frac{{k_y }}{k}}\right)^T~,~~
{\bf \tilde E}_{\rm slit}\!\propto\!\left({-\varepsilon,0,- \frac{{k_y }}{k}}\right)^{\!T}\!,~
\tilde{\bm{\mathcal E}}_p\!\propto\!\left({-i\chi,-i\chi\frac{k_y}{k_p},1}\right)^{\!T}\!,
\end{equation}
where $k_y /k \simeq \theta \sin{\phi}$. This results in the following coupling coefficient and the SPP field:
\begin{equation}\label{eqn:14}
\gamma \simeq -i\chi\varepsilon - \frac{k_y}{k},~
{\bf \tilde E}_p\!\propto\!\left({-i\chi\varepsilon - \frac{k_y}{k}}\right)\left({-i\chi,0,1}\right)^T.
\end{equation}
It is seen from here that the \textit{imaginary} longitudinal SPP field plays crucial role in the coupling with light. Indeed, the weak component of the input polarization, $\varepsilon$, is multiplied by $-i\chi=-i\kappa/k_p$ because of the coupling between the $x$-components of light and SPP fields. Therefore, \textit{imaginary} transverse wave number of SPPs, $k_{p_z} = -i\kappa$, effectively \textit{swaps} the real and imaginary parts of $\varepsilon$ in the coupling process. Hence, in order to make the weak-measurement formalism (1)--(7) consistent with our plasmonic system, one has to substitute $\varepsilon \to -i\chi\varepsilon$ or $\hat H_{\rm SOI} \to i\chi^{-1}\hat H_{\rm SOI}$ in Eqs.~(3)--(7). This modification immediately ascertains perfect agreement between the weak-measurement model and experimental results in Fig.~2.
%
%%%%%%%%%%%%%%%%%%%%%%%%%%%%%%%%%%%%%%%%%%%%%%%%%%%%%%%%%%%
\begin{figure}[t]
\includegraphics[width=8.5cm, keepaspectratio]{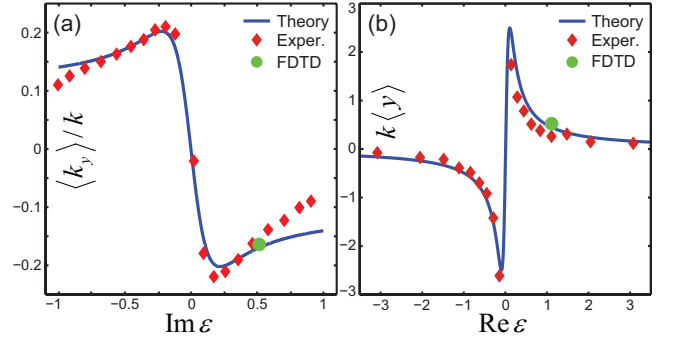}
\caption{(color online). Theoretically calculated (Eqs.~(15) with $\chi=1$ \cite{Remark II}) and experimentally measured SHEL shifts of the SPP beams versus complex pre-selection parameter $\varepsilon$ (see also \cite{SOM}). (a) -- imaginary $\varepsilon$ generating the momentum shift $\left\langle k_y \right\rangle$ and (b) -- real $\varepsilon$ producing the coordinate shift $\left\langle y \right\rangle$. For experimental convenience, the real space shifts were measured with NA=0.45 and Fourier-space shifts with NA=0.6.} \label{fig3}
\end{figure}
%%%%%%%%%%%%%%%%%%%%%%%%%%%%%%%%%%%%%%%%%%%%%%%%%%%%%%%%%%%
%

To compare experiment and theory quantitatively, we calculate the SPP beam centroids in the coordinate and momentum spaces. Taking into account that the position operator is $\hat y = i\partial /\partial k_y$ in the momentum representation, we define $\left\langle y \right\rangle  = \left\langle {\bf \tilde E}_p \right|\hat y\left| {\bf \tilde E}_p \right\rangle \left/ \left\langle {\bf \tilde E}_p \right|\left. {\bf \tilde E}_p \right\rangle \right.$ and $\left\langle k_y \right\rangle  = \left\langle {\bf \tilde E}_p \right| k_y \left| {\bf \tilde E}_p \right\rangle \left/ \left\langle {\bf \tilde E}_p \right|\left. {\bf \tilde E}_p \right\rangle \right.$, where the inner product implies the scalar product of the complex vector amplitudes and the integration in the momentum space: $\left\langle {{\bf \tilde E}_p } \right|\left. {{\bf \tilde E}_p } \right\rangle  \propto \int\limits_0^{2\pi }{\int\limits_0^{\theta_{\rm c}} {\left({{\bf \tilde E}_p^* \cdot {\bf \tilde E}_p} \right)}} \theta d\theta d\phi$. Performing these calculations with the SPP Fourier spectrum (14), we finally arrive at
\begin{equation}\label{eqn:15}
\left\langle y \right\rangle\!=\!\frac{1}{k\theta_{\rm c}}\,\frac{2\chi \theta_{\rm c}\,{\rm Re}\varepsilon}{2\chi^2|\varepsilon|^2 + \theta_{\rm c}^2/2},~
\left\langle{k_y}\right\rangle\!=\!-k\theta_{\rm c}\,\frac{\chi \theta_{\rm c}\,{\rm Im}\varepsilon}{2\chi^2|\varepsilon|^2 + \theta_{\rm c}^2/2}.
\end{equation}
We emphasize that Eqs.~(15) are valid in the \textit{whole range} of values of $\varepsilon$. In the weak-measurement range $\lambdabar/w_0 \ll \varepsilon \ll 1$, they are precisely equivalent to Eqs.~(7) with modification $\varepsilon \to -i\chi\varepsilon$, i.e., $\Delta \to i\chi^{-1} \Delta = \left({\chi k\varepsilon}\right)^{-1}$, and $w_0 = 2\left({k\theta_{\rm c}}\right)^{-1}$. Comparisons of the experimentally measured coordinate and momentum transverse shifts (as dependent on the complex polarization parameter $\varepsilon$) with the theoretical results (15) are shown in Figure~3. Moreover, we performed a finite difference time domain (FDTD) simulations for two input polarizations, and the resulting shifts are also presented in Fig.~3. Evidently, the experiment, wave theory, weak-measurement interpretation, and FDTD simulations are all in perfect agreement \cite{SOM}.

\textit{Conclusion.---}
%%%%%%%%%%%%%%%%%%%%%%%%%%%%%%%%%%%%%%%%%%%%%%%%%%%%%%%%%%%
We have observed and examined in detail an extraordinary plasmonic SHEL appearing in the interaction of focused light with a straight slit on the metal surface. Remarkably, this simple system offers a perfect weak-measurement tool where fixed polarization of SPPs provides a built-in post-selection. Tiny variations of the input polarization of light bring about huge SHEL shifts of the SPP beams in coordinate and momentum spaces, which correspond to the imaginary and real parts of the weak value of the helicity of light. The presented results demonstrate the unique ability of surface waves to perform as a post-selecting measuring device which might be potentially useful for various sensing applications involving the chirality of light.

We acknowledge fruitful discussions with Y. P. Bliokh and A. Y. Nikitin, and the financial support from the ERC (Grant 227577) and the European Commission (Marie Curie Action).

%\vspace*{-0.3cm}


\begin{thebibliography}{99}
\bibitem{Remark I} High-NA focusing also induces SOI and partial spin-to-orbital angular momentum conversion in the incident light \cite{Focusing1,Focusing2}. However, efficiency of this conversion is estimated as ${\rm NA}^2 /4 \simeq 0.05$ here and can be neglected against the strong SOI ($\propto{\rm NA}$) induced by the slit.
\end{thebibliography}
\end{document}